\documentclass[aps,prd,twocolumn,superscriptaddress,preprintnumbers,nofootinbib]{revtex4-1}
\usepackage{amsmath,amssymb,graphicx,natbib,bm}
\usepackage{enumerate}
\usepackage{hyperref}
\usepackage{breakurl}
\usepackage{color}

\newcommand{\fslash}[1]{\!\not\!{#1}}
\newcommand{\het}{\ensuremath{{}^3\mathrm{He}}}
\newcommand{\hef}{\ensuremath{{}^4\mathrm{He}}}

\def\be{\begin{equation}}
\def\ee{\end{equation}}
\def\ba{\begin{eqnarray}}
\def\ea{\end{eqnarray}}

\usepackage[table]{xcolor}
\usepackage[latin1]{inputenc}

\begin{document} 

\title{New physics via pion capture and simple nuclear reactions}

\author{Chien-Yi Chen}
\email{chien-yi.chen@northwestern.edu}
\affiliation{Department of Physics and Astronomy, University of Victoria, Victoria, BC V8P 5C2, Canada}
\affiliation{Perimeter Institute for Theoretical Physics, Waterloo, ON N2J 2W9, Canada}
\affiliation{Department of Physics and Astronomy, Northwestern University,  Evanston, Illinois 60208, USA}

\author{David McKeen}
\email{mckeen@triumf.ca}
\affiliation{TRIUMF, 4004 Wesbrook Mall, Vancouver, BC V6T 2A3, Canada}

\author{Maxim Pospelov}
\email{mpospelov@perimeterinstitute.ca}
\affiliation{Department of Physics and Astronomy, University of Victoria, Victoria, BC V8P 5C2, Canada}
\affiliation{Perimeter Institute for Theoretical Physics, Waterloo, ON N2J 2W9, Canada}

\begin{abstract}
Light, beyond-the-standard-model particles $X$ in the 1-100\,MeV mass range can be produced in nuclear and hadronic reactions but would have to decay 
electromagnetically. We show that simple and well-understood low-energy hadronic processes can be used as a tool to study $X$ production and decay. 
In particular, the pion capture process $\pi^- p \to X n \to  e^+  e^- n$ can be used in a new experimental setup to search for anomalies 
in the angular distribution of the electron-positron pair, which could signal the appearance of dark photons, axion-like particles and other exotic states. This 
process can be used to decisively test the hypothesis of a new particle produced in the $^7{\rm Li}+p$ reaction. We also discuss a variety of other 
theoretically clean hadronic processes, such as $p+{\rm D(T)}$ fusion, as a promising source of $X$ particles. 

\end{abstract}

\date{\today}

\maketitle

%%%%%%%%%%%%%%%%
%%%%%%%%%%%%%%%%

\section{Introduction}
\label{sec:intro}

Extensions of the Standard Model (SM) by light and weakly coupled particles such as axions, axion-like particles, dark photons, etc., 
have been recognized as a generic possibility. The particle physics community has evolved towards  
a very systematic approach to light beyond the SM (BSM) states, as several new experiments have been proposed and the results of 
old experiments re-analyzed~\cite{Battaglieri:2017aum,Beacham:2019nyx}. Some of the measurements have delivered ``anomalous'' results 
that may signify poorly understood SM physics, experimental problems, or indeed the existence of light, weakly coupled states. 
It is important to check the origin of such anomalies, and perhaps investigate the simplest BSM explanations, as was done in the case of the
muon $g-2$ discrepancy~\cite{Bennett:2006fi}, and its (as now excluded) solution via a dark photon~\cite{Pospelov:2008zw}. A similarly 
intense scrutiny (see, e.g.,~\cite{Barger:2010aj,*TuckerSmith:2010ra,*Batell:2011qq,*Carlson:2012pc,*Karshenboim:2014tka,*Liu:2016qwd}) followed the measurement of the muonic hydrogen Lamb shift, which seemed to be at odds with expectations using the 
charge radius of the proton determined by other means~\cite{Pohl:2010zza}. 

On the theoretical side, there is some better 
understanding of the allowed models of light particles. In particular for light vector $X$, the exact conservation of the SM current $X$ couples to 
(including cancellation of all anomalies) seems to be imperative for the construction of models that avoid 
$({\rm weak~scale})/m_X$ enhancement of the production amplitudes (see, e.g.,~\cite{Dror:2017ehi,Dror:2017nsg}). These theoretical rules 
could be ``bent'' at times (often signaling that some significant amount of fine-tuning needs to be tolerated), when a model can be considered as a candidate to 
solve an outstanding anomaly. This is exactly what has happened with the recently claimed anomaly in the angular distribution of $e^+e^-$ pairs produced in the 
proton capture by $^7$Li~\cite{Krasznahorkay:2015iga}. The experiment detected an unexpected dependence of pair yield on  the relative 
angle $\theta$ between the $e^+e^-$ pairs.  The anomaly manifests itself for the 18.15\,MeV intermediate state of $^8$Be, and is difficult 
to explain from the point of view of conventional nuclear physics approaches~\cite{Zhang:2017zap}. Instead, a hypothesis of 
an intermediate particle of mass $\simeq 17$\,MeV has been put forward~\cite{Krasznahorkay:2015iga}, that would modify the 
angular distribution in the desired way. The simplest models of vector $X$ coupled to quark currents have been suggested as an explanation~\cite{Feng:2016jff,Feng:2016ysn,Kahn:2016vjr,Kozaczuk:2016nma,Kozaczuk:2017per}, 
which all seem to require some tuning of parameters and/or physical amplitudes to avoid $m_X^{-1}$-enhanced processes~\cite{Dror:2017nsg}. 
Nevertheless, this so-called ``beryllium anomaly" is intriguing enough to study further. 

The goal of the present paper is as follows. We would like to argue that the simplest nucleon-involved processes can be used to search 
for light weakly coupled states, and check for the presence or absence of light resonances in the sub-20\,MeV region. {\it A priori}, the dynamics of 
eight nucleons inside $^8$Be can be quite complicated, and despite the study of~\cite{Zhang:2017zap}, one might not be able to currently conclude 
with all degree of certainty that the anomaly is not explained by conventional nuclear physics. On the other hand, processes that involve fewer 
 nucleons can be well understood, and the loophole of ``nuclear physics complication'' would not exist in such systems.  To that end, we would like to study the 
 simplest hadronic reactions, such as pion capture on protons as a possible source of $X$ particles. We would also like to point out other 
 promising reactions, such as proton fusion with deuterium or tritium as a source of exotic 17 MeV boson. In a less speculative vein, 
 we forecast sensitivity to  dark photons and other exotic bosons in a quasi-realistic pion capture set-up with attainable 
 pion intensities. 
 
 Previously, the $\pi^-p\to  e^+  e^- n$ reaction has been studied by the SINDRUM collaboration~\cite{MeijerDrees:1992kd},
 along with the pair production from $\pi^0$ produced in the charge-exchange reaction. The resulting constraints on dark photon 
 parameter space were derived in Ref.~\cite{Gninenko:2013sr}. Unfortunately, the experimental setup in~\cite{MeijerDrees:1992kd}
 did not allow probing pairs with invariant mass less than 20\,MeV, and therefore this past experiment cannot place constraints 
 on a hypothetical 17 MeV particle.

Pion capture on a proton is a powerful, robust way to search for the production of new, light, weakly coupled bosons. In the SM, capture of a negatively charged pion on a proton leads to the production of neutron which, to conserve momentum, recoils against a neutral boson with a mass less than about $m_{\pi^\pm}=139.6~\rm MeV$, either a photon or $\pi^0$. This should be contrasted with capture on nuclei. In that case, the capture process typically ejects a nucleon from the nucleus. The nucleon can then recoil against the nuclear remnant, thereby conserving momentum. As a result, fewer light, neutral bosons are released per event in $\pi^-$ capture on nuclei as compared to capture on a proton. In addition, capture on a proton has an additional advantage compared to that on a nucleus: less theoretical uncertainty in the calculation of rates because capture on a proton can be well understood in the context of a chiral effective Lagrangian. (Pion capture on 
deuterium can also be useful: while the yield of photons and new physics states are only a factor of $\sim2$ smaller than for capture 
on protons, the absence of $\pi^0$ in the final state may prove to be advantageous in reducing radiative backgrounds.)

New, weakly coupled bosons, $X$, with masses $m_X\lesssim m_{\pi^\pm}$ can also be produced in pion capture on a proton, $\pi^-p\to Xn$, if they couple to hadrons (or quarks). Production of these exotic bosons is relatively enhanced in $\pi^-$ capture on a proton as compared to nuclei for the same reason as $\pi^0$ and $\gamma$ production. Estimation of the rates of $X$ production are also theoretically cleaner in the nucleon system.

This paper is organized as follows. In the next section we review the radiative and pair-production pion capture in the SM. In Sec.~3 we provide relevant formulae for calculating the rate of $X$ production in the pion capture, where $X$ is either a
dark photon or ``protophobic" vector~\cite{Feng:2016jff,*Feng:2016ysn}. In the same section, we estimate the sensitivity  that can be achieved 
with modern sources of negative pions. In Sec.~4, we discuss additional nuclear physics channels (such as deuterium-proton fusion) 
that can be used in a new experiment looking for $X$. We reach our conclusion in Sec.~5, and generalize our results to axion-like 
particles in the Appendix.

%%%%%%%%%%%%%%%%
%%%%%%%%%%%%%%%%
\section{Radiative pion capture and pair production in the Standard Model}
\label{sec:calc}

Historically, the study of pion capture, the exothermic $\pi^-p$ reaction, was very important for understanding the 
simplest hadronic processes~\cite{Panofsky:1950he,Deser:1954vq}. 
To begin, we calculate the cross section for a $\pi^-$ bound to a proton to produce a photon and a neutron. We will use the prediction of this rate
(and corresponding experimental measurement)  to normalize the production rates of exotic states later on.

\begin{figure}
\includegraphics[width=\linewidth]{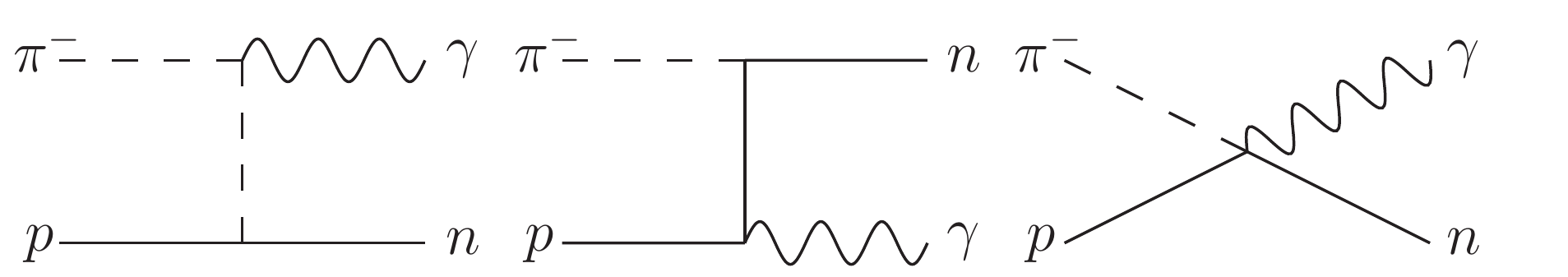}
\caption{Diagrams responsible for the reaction $\pi^-p\to\gamma n$ from the interactions in Eq.~(\ref{eq:chiralEFT}).}\label{fig:ngamma}
\end{figure}

This process is described well in the low-energy chiral Lagrangian involving nucleons and pions. The relevant terms for this process read
\begin{equation}
\begin{aligned}
{\cal L}&\supset \left|\left(\partial_\mu+ieA_\mu\right) \pi^-\right|^2-m_\pi^2\pi^+\pi^-
\\
&+\bar p\left(i\!\fslash{\partial}+e\!\fslash{\!A}-m_N\right)p+\bar n\left(i\!\fslash{\partial}-m_N\right)n
\\
&-\frac{g_A}{\sqrt2 F_\pi}\bar n\gamma^\mu\gamma^5 p \left(\partial_\mu+ieA_\mu\right) \pi^-,
\label{eq:chiralEFT}
\end{aligned}
\end{equation}
where $g_A=1.275$~\cite{Czarnecki:2018okw} is the nucleon axial vector coupling and $F_\pi=92~\rm MeV$~\cite{Agashe:2014kda} is the pion decay constant. We do not differentiate between proton and neutron mass, neglect particle form-factors and anomalous magnetic moment contributions.  
There are three separate amplitudes, shown in Fig.~\ref{fig:ngamma}, that contribute to the process $\pi^-p\to\gamma n$.

A straightforward calculation of the cross section for this process gives
\begin{equation}
\begin{aligned}
\left(\sigma v\right)_{\pi^-p\to\gamma n}&=\frac{\alpha g_A^2}{F_\pi^2}\frac{1+m_\pi/(2m_N)}{1+m_\pi/m_N}
\\
&=5.1\times10^{-28}~{\rm cm}^2\times c. 
\label{eq:estimate}
\end{aligned}
\end{equation}

The branching fraction for radiative pion capture can be computed using the measured value of the  Panofsky ratio~\cite{Spuller:1977ve},
\begin{equation}
P=\frac{\left(\sigma v\right)_{\pi^-p\to\pi^0 n}}{\left(\sigma v\right)_{\pi^-p\to\gamma n}}=1.546\pm0.009,
\end{equation}
which relates the strength of this mode to the other final state, $\pi^0 n$. From this one finds a radiative branching
\begin{equation}
\label{brAn}
{\rm Br}_{\pi^-p\to\gamma n}=\frac{1}{1+P}=0.3928\pm0.014.
\end{equation}
Calculation of the rate can be contrasted with the measurement of the $1s$ width~\cite{Gotta2008}, which can be converted to the capture rate, 
\begin{equation}
\begin{aligned}
\left(\sigma v\right)_{\pi^-p\to\gamma n}^{\rm meas}& =\Gamma_{1s}^{\rm meas}{\rm Br}_{\pi^-p\to\gamma n} \times |\psi_{1s}(0)|^{-2}
\\
&\simeq 5.7 \times 10^{-28}~{\rm cm}^2\times c,
\end{aligned}
\end{equation}
with $\sim 3\%$ experimental error. The accuracy of our theoretical result in Eq.~(\ref{eq:estimate}) is within $\sim 10\%$ of this measurement, which is rather good 
given crude nature of some of the approximations. This accuracy is entirely sufficient for our purposes of studying new physics. 

We also calculate the rate for the $\pi^-p\to e^+e^-n$ process, given the Lagrangian of Eq. (\ref{eq:chiralEFT}). 
The result, in terms of the invariant mass of a pair, $m_{ee}$, for $m_e/m_\pi \ll1$ is given by
\begin{equation}
\frac{d\left(\sigma v\right)_{\pi^-p\to e^+e^-n}}{\left(\sigma v\right)_{\pi^-p\to\gamma n}} = \frac{2\alpha}{3\pi} \frac{dm_{ee}}{m_{ee}}\times 
f\left(\frac{m_{ee}}{m_N},\frac{m_\pi}{m_N}\right)
\label{eq:bkgdist}
\end{equation}
with
\begin{equation}
\begin{aligned}
&f(x,y)=\sqrt{1-\frac{x^2}{y^2}}\frac{\left(1-z^2\right)^{3/2}}{\left(1-xz\right)^{2}\left(1-xz/y^2\right)^{2}}
\\
&\!\times\left\{1-\frac{z^2}{2y^2}\left[7-4y^2-\frac{x^2}{y^2}\left(2-2y-4y^2+x^2\right)\right]\right\}
\end{aligned}
\label{eq:f}
\end{equation}
and $z=x/(2+y)$.
This function is defined so that $f\to 1$ for $m_{ee} \ll m_\pi$. The logarithmic term, $ \propto dm_{ee}/m_{ee}$, is perfectly 
consistent with the classic paper by Kroll and Wada~\cite{Kroll:1955zu}. The remaining integral can be performed to find the full probability of emitting a pair
relative to emitting a photon, which is dominated by small $m_{ee}$ and is about $7.3\times 10^{-3}$. In practice, the relative angle between final state charged particles, rather than their invariant mass, represent a more convenient variable. 

\section{Production of new bosons in pion capture}

Having calculated the rate for radiative $\pi^-$ capture on a proton, we would now like to estimate the rates to emit new, weakly coupled bosons, $X$, in the same process. We begin with the well-studied ``dark photon,'' $X=A^\prime$. The $A^\prime$ is a vector boson with mass $m_{A^\prime}$ that kinetically mixes with the photon with strength $\epsilon$. Its couplings to SM particles are then the same as the photon's times the mixing strength. We can therefore use Eq.~(\ref{eq:chiralEFT} with the replacement $A_\mu\to\epsilon A^\prime_\mu$ to describe the interactions of the dark photon with hadrons. (We do not 
consider the $m_{A'}\to 0$ limit, where dark photon interactions are additionally suppressed.) There are again three amplitudes that contribute, completely analogous to those in Fig.~\ref{fig:ngamma}. The rate for dark photon production, $\pi^-p\to A^\prime n$, relative to that of standard photons is then
\begin{equation}
\label{eq:dph} 
\begin{aligned}
&\frac{\left(\sigma v\right)_{\pi^-p\to A^\prime n}}{\left(\sigma v\right)_{\pi^-p\to\gamma n}}=\epsilon^2f\left(\frac{m_{A^\prime}}{m_N},\frac{m_\pi}{m_N}\right),
\end{aligned}
\end{equation}
with $f$ from Eq.~(\ref{eq:f}) describing the phase space dependence of the process. 

\begin{figure}
\includegraphics[width=\linewidth]{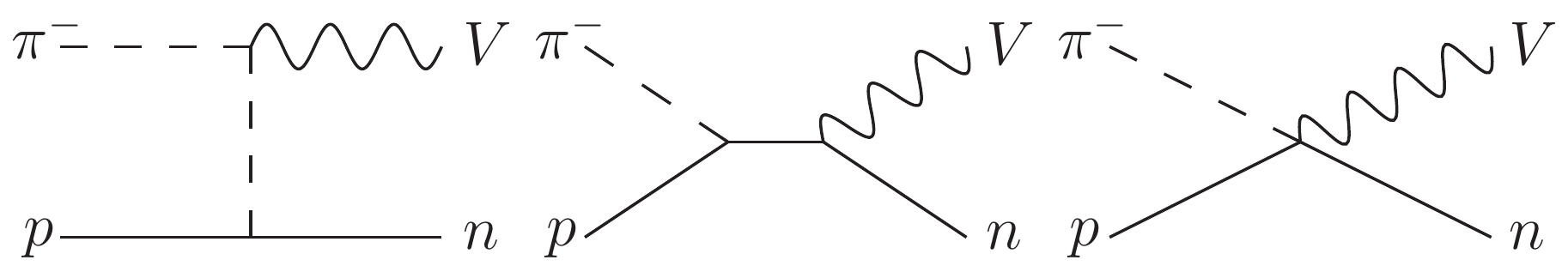}
\caption{Diagrams responsible for the reaction $\pi^-p\to\gamma n$ from the interactions in Eq.~(\ref{eq:chiralEFT}).}\label{fig:nV}
\end{figure}

\begin{figure}
\includegraphics[width=\linewidth]{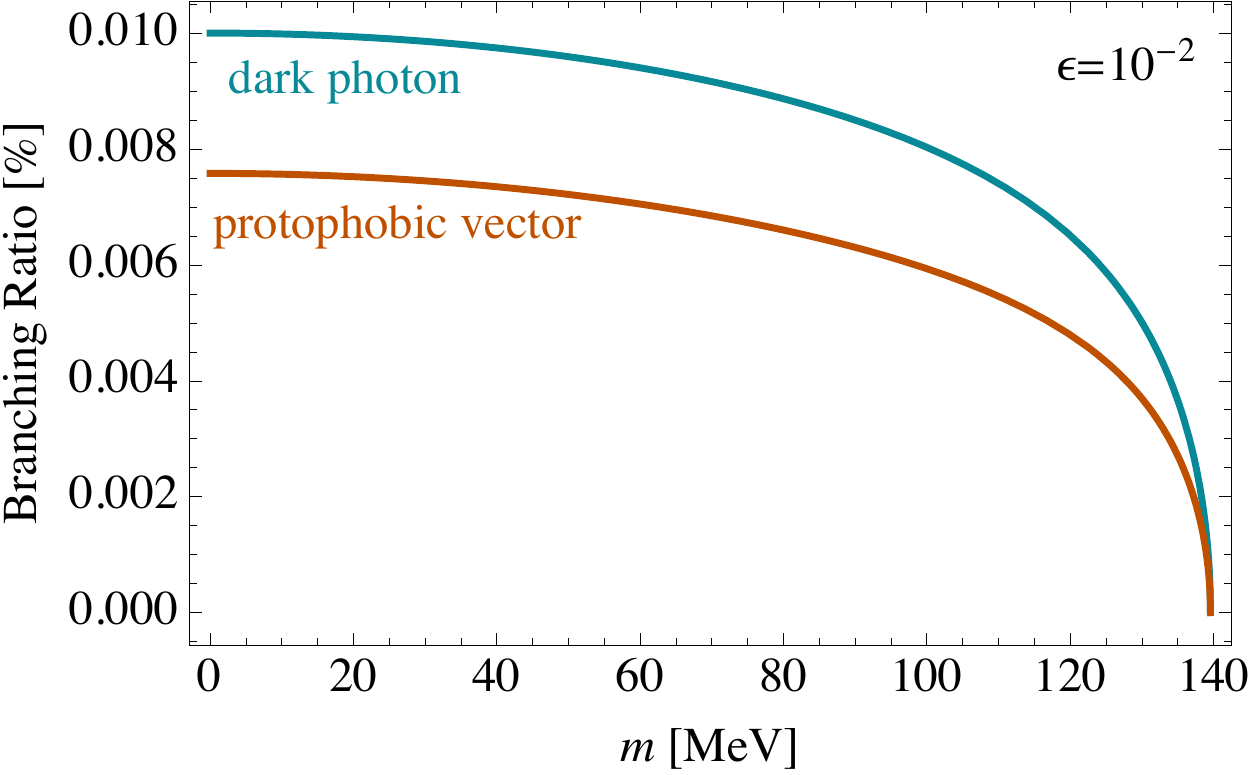}
\caption{The branching ratios for $\pi^- p\to A^\prime n$ (dark photon) and $\pi^- p\to V n$ (protophobic vector) as functions of the vector masses. We have taken $\epsilon=10^{-2}$ in both cases.}\label{fig:branchings}
\end{figure}

As another benchmark, we take the recently proposed ``protophobic'' vector boson~\cite{Feng:2016jff,*Feng:2016ysn} that can explain the excess $e^+e^-$ events seen in nuclear transitions of $^8{\rm Be}$~\cite{Krasznahorkay:2015iga}. In this setup, a new vector boson $X=V$, with a mass $m_V\simeq17~\rm MeV$ is introduced that couples with strength $\epsilon\times e$ (choosing this normalization to make contact with the dark photon case) to neutrons and negative pions but not to protons. Note that the coupling of $V$ to $n$ and $\pi^-$ has the same sign. In addition, $V$ has a separate coupling to electrons that mediates its prompt decay to $e^+e^-$.

To calculate the rate of $V$ production in pion capture, we need its couplings to hadrons. These can be easily lifted from the Lagrangian of Eq.~(\ref{eq:chiralEFT}) with suitable adjustments of the covariant derivatives. The resulting interactions are described by
\begin{equation}
\begin{aligned}
{\cal L}&\supset \left|\left(\partial_\mu-i\epsilon eV_\mu\right) \pi^-\right|^2-m_\pi^2\pi^+\pi^-
\\
&+\bar p\left(i\!\fslash{\partial}-m_N\right)p+\bar n\left(i\!\fslash{\partial}+\epsilon e\fslash{V}-m_N\right)n
\\
&-\frac{g_A}{\sqrt2 F_\pi}\bar n\gamma^\mu\gamma^5 p \left(\partial_\mu-i\epsilon eV_\mu\right) \pi^-
\end{aligned}
\label{eq:protophobL}
\end{equation}

As shown in Fig.~\ref{fig:nV}, the amplitude to produce a photophobic vector, $\pi^-p\to V n$, contains three terms, as in the (dark) photon case. However, in this case, instead of emission off of a proton, one term involves emission from the neutron. Adding these contributions together, the capture 
rate with the photophobic vector production is found to be
\begin{equation}
\label{eq:dph2}
\frac{\left(\sigma v\right)_{\pi^-p\to V n}}{\left(\sigma v\right)_{\pi^-p\to\gamma n}}=\frac{\epsilon^2}{\left(1+m_\pi/m_N\right)^2}\,g\left(\frac{m_V}{m_N},\frac{m_\pi}{m_N}\right),
\end{equation}
where the phase space function is given by
\begin{equation}
\begin{aligned}
g(x,y)&=\sqrt{1-\frac{x^2}{y^2}}\left[1-\left(\frac{x}{2+y}\right)^2\right]^{3/2}
\\
&\quad\times\left\{1+\frac12\left[\frac{xy\left(1+y\right)}{\left(2+y\right)y^2-x^2}\right]^2\right\}.
\end{aligned}
\end{equation}

The branching ratios for $\pi^- p\to A^\prime n$ (dark photon) and $\pi^- p\to V n$ (protophobic vector) can be obtained by multiplying ${\rm Br}_{\pi^-p\to\gamma n}$ in Eq.~\ref{brAn} by Eqs.~\ref{eq:dph} and ~\ref{eq:dph2}, respectively. We show the branching ratios  as functions of the vector masses in Fig.~\ref{fig:branchings}, where we have taken $\epsilon=10^{-2}$ in both cases.  

\section{Experimental Concept}
\label{sec:exp}
The signature of a new boson that decays to $e^+e^-$  produced in $\pi^-$ capture (both the dark photon and protophobic vector considered above) is the production of an $e^+e^-$ pair with an invariant mass peaked at $m_{ee}=m_X$. Since the sum of the electron and positron energies is fixed to be 
close to $m_\pi$, a peak in the invariant mass also translates into a sharp feature in the opening angle of the pair. A hypothetical particle invoked 
as an explanation of the anomaly in $^8$Be$^*$ decay would be seen in pion capture as a sudden increase of lepton pairs with opening angles
larger than $\sim 14$ degrees.  

This signal sits on top of Standard Model processes leading to a pair-creation. 
In particular, Dalitz decays of final state $\pi^0$, $\pi^0\to \gamma e^+e^-$ would constitute a significant source of background, which can 
however be removed by requiring $E_{e^+}+E_{e^-}>m_\pi/2$. Another source of background, the photon conversion in material, 
 following the radiative capture can also be controlled using experimental means and requiring a symmetric distribution of $e^+$ and $e^-$ energies. There is, however, one source of 
an irreducible SM background from pion capture that produces an off-shell photon,  $\pi^-p\to\gamma^\ast n\to e^+e^-n$. Because of the photon pole in the process, the SM invariant mass distribution is a monotonically decreasing function of $m_{ee}$, given in Eq. (\ref{eq:bkgdist}).  

Ignoring the electron mass, $m_{ee}^2\simeq 2E_+E_-(1-\cos\theta)$ with $E_+$ and $E_-$ the energies of the $e^+$ and $e^-$ respectively and $\theta$ the opening angle between them. Therefore, the precision reconstructing $m_{ee}$ is affected by the finite energy and angular resolution of any experimental setup. We can estimate the reach of a ``bump hunt'' in $m_{ee}$ by considering the number of signal events in a bin of width $\delta_{ee}$ centered on $m_{ee}=m_X$ which is roughly
\begin{equation}
N_{\rm sig}\sim\epsilon^2 N_{\rm cap}{\rm Br}_{\pi^-p\to\gamma n} ,
\end{equation}
with $N_{\rm cap}$ the number of $\pi^-$ captures and $\epsilon$ the $X$ coupling in units of the positron charge. The size of the bin $\delta_{ee}$ is determined by the experimental resolution that can be achieved for $E_\pm$ and $\theta$. The number of background events in the bin centered around $m_X$ can be found using Eq.~(\ref{eq:bkgdist})
\begin{equation}
N_{\rm bkg}\sim \frac{2\alpha}{3\pi}\frac{\delta_{ee}}{m_X}N_{\rm cap}{\rm Br}_{\pi^-p\to\gamma n} ,
\end{equation}
where we have ignored terms of ${\cal O}(\delta_{ee}^2/m_X^2)$. 

Requiring that the number of signal events in this bin is larger than a $3\sigma$ statistical fluctuation of the background, one can estimate the values of $\epsilon$ that can be probed, which scales as $(N_{\rm cap})^{-1/4}$,
\begin{equation}
\begin{aligned}
\epsilon&\gtrsim 0.02\left(\frac{\delta_{ee}/m_X}{30\%}\right)^{1/4}\times \left(\frac{10^4}{N_{\rm cap}}\right)^{1/4}, 
\\
N_{\rm sig}&\gtrsim3 N_{\rm bkg}^{1/2}
\end{aligned}
\end{equation}
The size of the bin near $m_{ee}=m_X$, $\delta_{ee}$, is important in controlling the sensitivity to $X$ production. This is determined by the experimental resolution on $m_{ee}$ which depends on the resolution measuring the $e^\pm$ energies and opening angle. A well known difficulty arises from the following: $E_+$ and $E_-$ are unlikely to be measured 
very precisely using calorimetric tools. But if the measurement of momentum via tracking of charged particles in the magnetic field 
is employed, then the rescattering of leptons in the tracking layers lead to the significant broadening of $\theta$. Large values of 
$\delta_{ee}$ would necessarily bring large number of the background events in the corresponding bin. 

  In practice, obtaining large rates while keeping the energy resolution small enough to have good precision on $m_{ee}$ is a challenge. 
  (For example, a $\sim 30\%$ resolution in energy, would result in $\delta_{ee}/m_X \sim 15\%$ even with a perfect measurement of $\theta$.)
   In contrast, the angular resolution can typically be kept under control; for this reason $\theta$ can be a useful discriminant for distinguishing signal from irreducible background, as taken advantage of in~\cite{Krasznahorkay:2015iga}.

For  $e^+e^-$ production through an on-shell boson $X\to e^+e^-$ produced in pion capture, there is a lower bound on the opening angle at
\begin{equation}
\theta_{\rm min}=\cos^{-1}\left(1-2m_{X}^2/m_\pi^2\right)=\frac{2m_{X}}{m_\pi}+{\cal O}\left(\frac{m_{X}^3}{m_\pi^3}\right).
\end{equation}
The background through an off-shell photon has no such cutoff and is peaked toward $\theta=0$. The intrinsic broadening of $\theta$ may come from 
various sources. The uncertain position of pion capture within a target leads to a geometric uncertainty in $\theta$. The thickness of the target 
(and in case of a cryogenic H$_2$ target, the thickness of walls around the target) contributes to multiple Coulomb scattering. We estimate that 
an electron and positron pair with individual energies $\sim m_\pi/2$ will broaden $\theta$ by $\Delta\theta\sim 0.5$ degree after 10 cm of liquid 
H$_2$ (such target dimensions are sufficient to stop $\pi^-$ with $p_\pi \leq 70$\,MeV). If a detector registering leptons is placed $1~{\rm m}$ away from the target, the geometric broadening is small. Thus, we believe that the resolution of $1^\circ$ on $\theta$ can be achieved. 

\begin{figure}
\includegraphics[width=\linewidth]{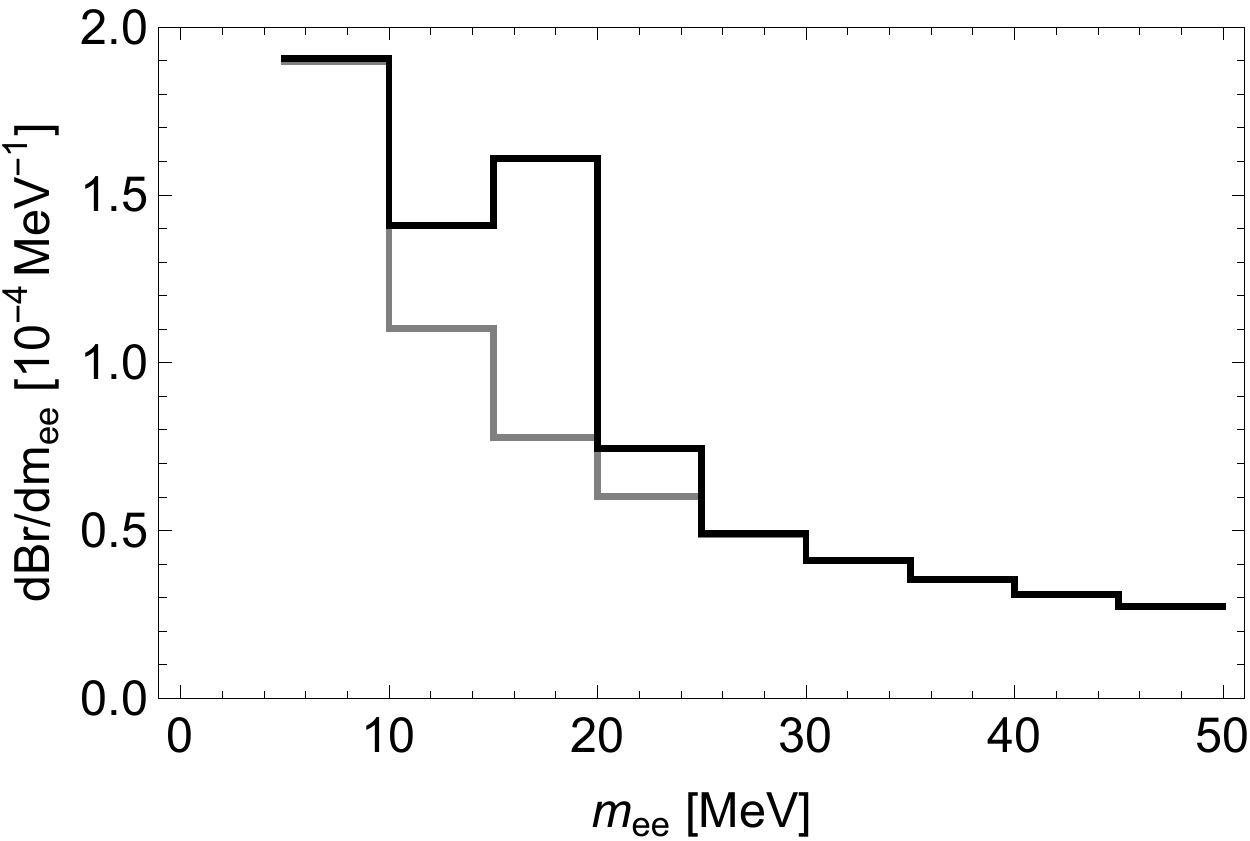} \\
\includegraphics[width=\linewidth]{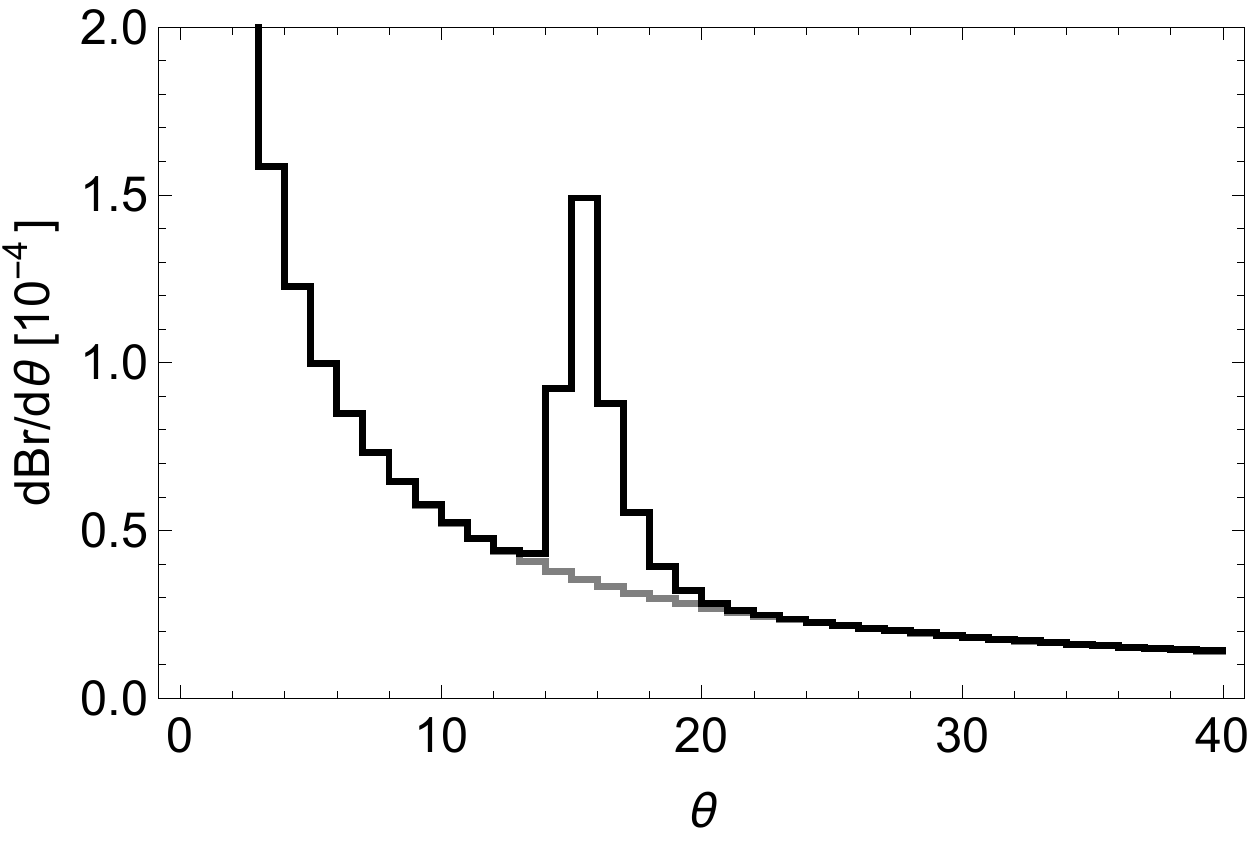}
\caption{The distribution of $e^+e^-$ invariant mass (top) and opening angle (bottom) in $\pi^-$ capture on a proton due to both background through an off-shell photon (gray) and background plus the signal of a protophobic vector (black), cf. Eq.~(\ref{eq:protophobL}). We take $m_V=17~{\rm MeV}$ as suggested by the $^8{\rm Be}$ anomaly and $\epsilon=0.1$. We assume a $1^\circ$ resolution on the $e^\pm$ direction and a $50\%$ resolution on their energies.}\label{fig:dists}
\end{figure}

To understand the potential sensitivity of a pion capture experiment more qualitatively, we simulate both background and signal from a protophobic vector in~(\ref{eq:protophobL}) decaying to $e^+e^-$ with $m_V=17~\rm MeV$. We show the resulting $m_{ee}$ distribution with $\epsilon=0.1$ assuming a $50\%$ resolution on the $e^+$ and $e^-$ energies and a momentum direction resolution of $1^\circ$ in Fig.~\ref{fig:dists}. We impose a cut $E_++E_->m_\pi/2$ to reduce the background from Dalitz decays of $\pi^0$'s.
We also show the opening angle distribution of both signal and background in Fig.~\ref{fig:dists}, requiring that
\begin{equation}
-0.5\leq\frac{E_+-E_-}{E_++E_-}\leq 0.5
\end{equation}
to enhance the signal to background ratio, since the signal events have a more symmetric distribution of energies.

\begin{figure}
\includegraphics[width=\linewidth]{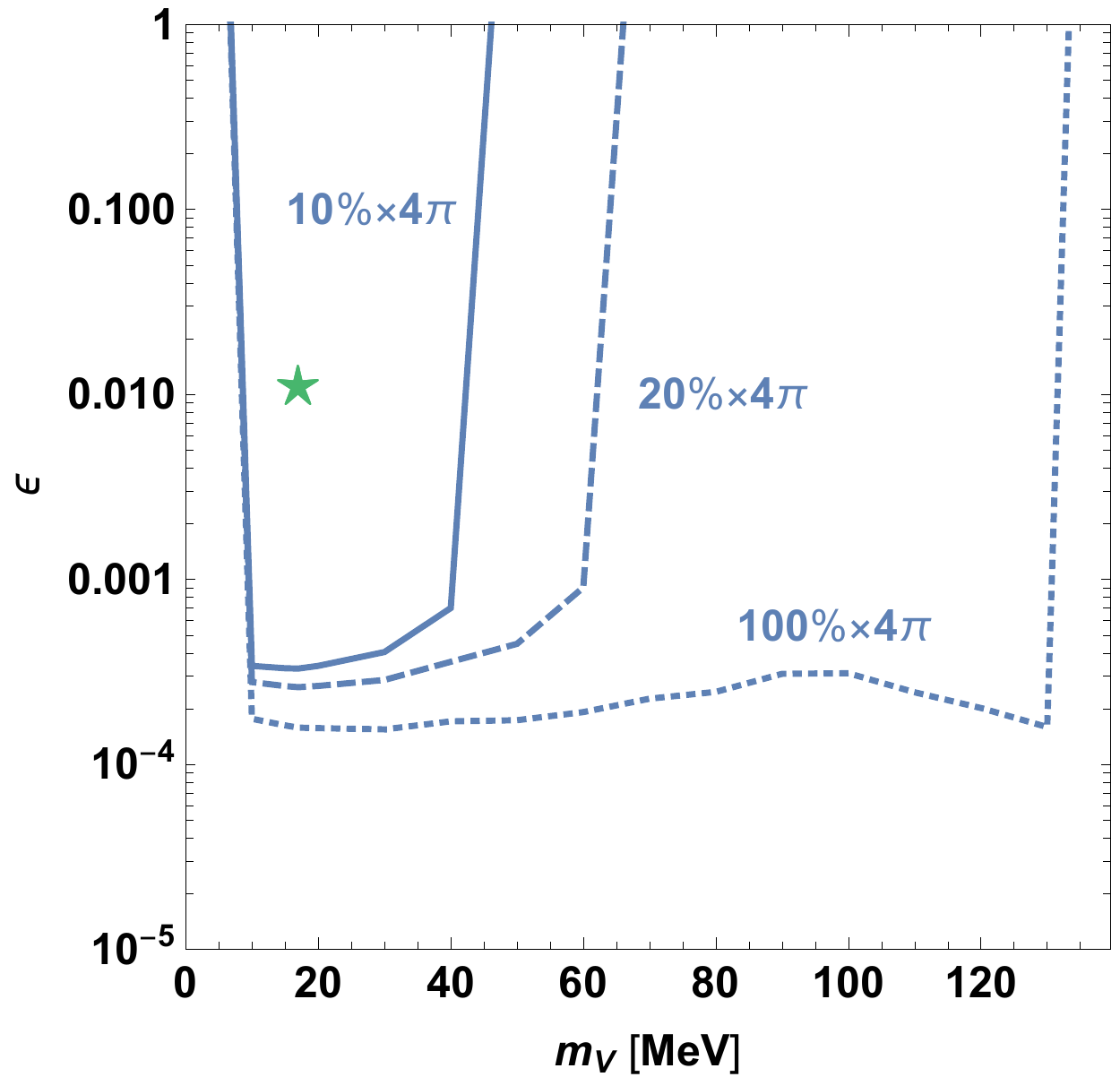}
\caption{Minimum values of the coupling of a protophobic boson that can be probed through production in $\pi^-$ capture on a proton for $10^{10}$ captures, an $e^\pm$ energy resolution of $50\%$ and $1^\circ$ resolution on the direction of the $e^\pm$ momenta. For each mass point we require the reconstructed $e^+e^-$ opening angle is within $[-1^\circ,+4^\circ]$ of $\theta_{\rm min}(m_V)$ and $-0.5\leq(E_+-E_-)/(E_++E_-)\leq 0.5$. The reach is determined by requiring the number of signal events is larger than a $3\sigma$ statistical fluctuation of the background. The solid, dashed, and dotted blue lines show the reach for detectors that have $10\%$, $20\%$, and $100\%$ coverage of the full $4\pi$ solid angle, respectively. The green dot shows the coupling and mass able to explain the $^8{\rm Be}$ anomaly~\cite{Feng:2016jff,*Feng:2016ysn}.}\label{fig:reach}
\end{figure}

We use this simulation to estimate the minimum values of the coupling of a protophobic boson that can be probed through production in $\pi^-$ capture on a proton for $10^{10}$ captures (corresponding to $\sim 1$ day of running at PSI). We require for each choice of $m_V$ that the $e^+e^-$ opening angle satisfies $\theta-\theta_{\rm min}(m_V)\in[-1^\circ,+4^\circ]$. We further enforce $E_++E_->m_\pi/2$ and $-0.5\leq(E_+-E_-)/(E_++E_-)\leq 0.5$ to cut down backgrounds as described above and assume $50\%$ and $1^\circ$ energy and angular resolutions, respectively. The results of this procedure are shown in Fig.~(\ref{fig:reach}). This minimum value of $\epsilon$ that can be probed is determined by the requirement that the number of signal events with this selection is larger than a $3\sigma$ statistical fluctuation of the background. We show the reach for detectors that have $10\%$, $20\%$, and $100\%$ coverage of the full $4\pi$ solid angle along with the parameters that explain the $^8{\rm Be}$ anomaly~\cite{Feng:2016jff,*Feng:2016ysn}. (A required value of $\epsilon^2$ in the protophobic case can be estimated via $v^{-3}_V\times 
(\Gamma_V/\Gamma_\gamma)$, where $v_V\simeq0.35$ is the final state velocity of the suggested 17 MeV particle, and ratio of exotic-to-gamma emission $\Gamma_V/\Gamma_\gamma$ in the decay of the 18.15\,MeV state is the claimed experimental value of $5.8\times 10^{-6}$ \cite{Krasznahorkay:2015iga}.)
Clearly, larger boson masses require a detector that can cover larger solid angles since they move with smaller velocity in the lab frame so that the $e^+e^-$ pair is less boosted and emitted closer to back-to-back.

%%%%%%%%%%%%%%%%%%%%%%%%%%%%%%%%%%%%

\section{Exotic bosons in nuclear reactions}

Pion capture represents perhaps the simplest hadronic reaction where a new boson can be explored or constrained without much complication 
of nuclear physics. At the same time, powerful sources of $\pi^-$ (with sub-100 MeV momentum so that they can be stopped) exist only in a handful of laboratories around the world. It is then important to consider other very simple hadronic reactions, that involve up to $4$ nucleons, and where 
nuclear physics may also be considered ``under control."

One can discuss the following as candidate radiative reactions,
\begin{eqnarray}
p+n &\to & {\rm D} +\gamma(X), ~~Q=2.2\,{\rm MeV},\\
\label{Dp}
{\rm D}+p &\to & \het +\gamma(X), ~~Q=5.5\,{\rm MeV},\\
{\rm D}+n &\to & {\rm T} +\gamma(X), ~~Q=8.5\,{\rm MeV},\\
{\rm T}+p &\to &\hef +\gamma(X), ~~Q=19.8\,{\rm MeV},\\
\het+n &\to & \hef +\gamma(X), ~~Q=20.6\,{\rm MeV},
\end{eqnarray}
where instead of a $\gamma$ one could emit an exotic boson $X$. The last two reactions have enough energy to 
search for a 17\,MeV state even for a small energy of incoming particles, while the first three reactions would require
somewhat energetic beam of protons or neutrons. 

On the other hand, if a beam of energetic protons up to a few tens of MeV is available, 
the ${\rm D}+p \to \het +\gamma(X)$ reaction is an ideal testing ground for the search of 
exotic bosons. The protons provide a powerful tool for such a search due to the potentially much larger statistics 
relative to pion capture. In addition, this process is in principle calculable to a great degree of accuracy \cite{Marcucci:2005zc}, 
so that the complications of nuclear theory are not an issue. Moreover, pair production via 
${\rm D}+p \to \het +e^+e^-$ has been successfully studied experimentally in the past~\cite{Johansson:1998jc}. 

It is easy to relate the energy of a proton beam, $E_p$, in the lab frame to the maximum available energy in the center-of-mass frame
for the emission of $X$, 
\begin{equation}
E_{\rm max} = E_p \frac{m_{\rm D}}{m_{\rm D}+m_p} + Q  = \frac{2}{3} E_p +Q. 
\label{Emax}
\end{equation}
For example, a proton of energy 20 MeV capturing on $\rm D$ corresponds to $E_{\rm max} = 18.8$\,MeV, which is enough to test the hypothesis of 
a light particle of $m_X = 17$\,MeV mass.

 Assuming the dominance of the $E1$ amplitude we can relate the emission of a vector boson $X$ to the emission of 
 $\gamma$ in the ${\rm D}(p,\gamma)\het$ reaction. We start with the dark photon, $X=A^\prime$. 
 In the assumption that the matrix elements of longitudinal current are the same as for the transverse current (see, e.g.,~\cite{Kroll:1955zu}
 for the details of such separation, and recent calculations of $e^+e^-$ emission from nuclear reactions in \cite{Pitrou:2019pqh}), 
 the rate is rather simply related to the photon rate, 
 \begin{equation}
 \label{eq:Dp}
 \sigma_{{\rm D}+p\to \het +A'} = \epsilon^2  \sigma_{{\rm D}+p\to\het +\gamma} \times \frac{v_{A'}(3-v_{A'}^2)}{2}.
 \end{equation}
 Here $v_{A'} = (1-m_{A'}^2/E_{max}^2)^{1/2}$ is the velocity of the outgoing $A'$ particle in the center-of-mass frame. 
 In the limit of the small mass, the rate into dark photons is simply $\epsilon^2$ from the rate to the regular photons. 
Going through the calculation of the relative dipole for the ${\rm D}-p$ system, one can see that in the protophobic case 
the size of the dipole is the same as in the standard ${\rm D}+p\to \het +\gamma$ reaction, multiplied by $\epsilon$. 
Therefore, the very same formula (\ref{eq:Dp}) would apply to the protophobic case as well. We note that 
the rate of dark photon emission can be related to the SM ${\rm D}+p \to \het +e^+e^-$ rate bypassing any theoretical assumptions, 
as for $m_{ee}= m_{A'}$ the emission of the pair and the emission of dark photon has the very same kinematic dependence on all
relevant  energies, cf. Eqs. (\ref{eq:bkgdist}) and (\ref{eq:dph}),
\begin{equation}
\sigma_{{\rm D}+p\to \het +A'} = \left.\epsilon^2 \frac{d\sigma_{{\rm D}+p\to \het +e^+e^-}/dm_{ee}}{2\alpha/(3\pi m_{ee})}\right|_{m_{ee}=m_{A'}}.
\end{equation}

It is easy to see that for realistic proton beams ($E_p\sim 20-40$\,MeV, up to $1~{\rm mA}$ currents) and gas ${\rm D}_2$ targets, one can achieve 
production of $V$-bosons at rates as high as $10^4\,{\rm s}^{-1}$ for $\epsilon^2=10^{-4}$. Therefore, the main difficulty in 
such an experiment would not be low statistics but strong backgrounds from the elastic and inelastic scattering of $p$ on ${\rm D}$. 
This could be circumvented if the final state \het\ is detected in coincidence with electrons and positrons (see, e.g.,~\cite{Adlarson:2014ysb}). 
Therefore, we believe that the simplest nuclear reactions, including the ones that require energetic protons, 
can also be used as a tool for studying $X$-bosons in realistic experiments, without much complication coming from nuclear physics 
as in the case of $^8$Be$^*$.

\section{Conclusions}
\label{sec:conclusions}

Simple low-energy hadronic processes may provide an efficient way of probing light new physics. 
The simplest process, $\pi^-$ capture on protons, can be used to search for light degrees of freedom coupled 
to quarks (i.e. to pions and nucleons at low energy), and decaying electromagnetically. Our analysis shows that 
with a powerful source of sub-100 MeV $\pi^-$, 
one can probe the dark photon parameter space down to $\epsilon \sim 10^{-3} $ in the mass range of a few MeV to $m_\pi$.
We note that only a very few experiments are sensitive in the $10-30$\,MeV mass range, where higher energy probes 
(e.g., $B$-factories) become less efficient. 

We have also considered the production of the so-called ``protophobic'' dark vector $V$ that was suggested~\cite{Feng:2016jff,Feng:2016ysn}
as a candidate explanation for the unusual angular correlation observed in the decay of a highly excited state of $^8$Be~\cite{Krasznahorkay:2015iga}.
In a hypothesis that the angular anomaly stems from a new $X$ particle, pion capture should exhibit similar unexplained variation of 
distribution in the relative electron-positron angle at around 14 degrees. The ``protophobic'' dark vector hypothesis can be then decisively tested this way, 
free of any possible nuclear physics complications. 

Another possible avenue for searches of $X$ is given by nuclear reactions in few-nucleon systems. 
In particular, the previously studied ${\rm D}+p \to \het +e^+e^-$ process may be used for $X$ searches.  With proton beams in excess of 20 MeV,
and with the same coupling size to explain the beryllium anomaly, copious production of $X$ bosons can be achieved, potentially opening a new avenue for their study.

\subsubsection*{\bf Acknowledgements}
We thank Drs. D. Bryman, J. Feng and A. Papa  for helpful discussions and useful communications. MP is grateful 
to Drs. B. Bastin, A. Coc and his colleagues at CSNSM Orsay, for productive discussions of possible $X$-boson search strategies.  
Research at Perimeter Institute is supported by the Government of Canada through Industry Canada and
by the Province of Ontario through the Ministry of Economic Development \& Innovation. C.-Y.C is supported by the DOE grant DE-FG02-91ER40684 and the NSF grant NSF-1740142. DM is supported by a Discovery Grant from the Natural Sciences and Engineering Research Council of Canada (NSERC) and TRIUMF which receives funding through a contribution agreement with the National Research Council of Canada (NRC).

\appendix

\section{New physics models}

In this appendix, we will discuss some potential models where the new boson is a pseudoscalar or an axial vector. Each of them can be in the flavor SU$(2)$ singlet or triplet representation. As with the vector bosons we discussed in the text, such states can be produced in simple hadronic reactions, such as pion capture on a proton. We give expressions for this capture rate to facilitate the easy computation of sensitivities.

\subsection{Singlet and triplet pseudoscalars}

The Lagrangian for a singlet pseudoscalar ($a$) coupling to the axial current $j_5^\mu\equiv {\bar {N}} \gamma^\mu \gamma_5 N$ can be written as
\be
{\cal L} \supset {g_a^S \over F_a} \partial_\mu a {\bar {N}} \gamma^\mu \gamma_5 N 
+ {g_a^{S'} \over F_a F_\pi} \pi^b a \; {\bar {N}} \tau^b N.
\ee
where $\tau^b$ for $b= 1$ to 3, are the Pauli matrices and $N=(p \;\, n)^{T}$ is a $SU(2)$ doublet with $p$ and $n$ the proton and neutron, respectively. 
The second term is analogous to the effective operator $1/(2 F_\pi^2) \pi^2  {\bar N} N \sigma$, in the chiral perturbation
theory, where $\sigma = {m_u+m_d \over 2}\langle N|\bar{u}u+\bar{d}d|N\rangle $ and
$g_a^S=\Delta u +\Delta d \simeq 0.521$ \cite{Barone:2001sp,Bishara:2016hek} with $2 s^\mu \Delta u=\langle p|\bar{u}\gamma^\mu\gamma_5 u |p\rangle $ and $2 s^\mu \Delta d=\langle p|\bar{d}\gamma^\mu\gamma_5 d|p\rangle $, where  $s^\mu$ is the proton spin vector.
From the fact that \cite{Borsanyi:2014jba}
\be
{m_u-m_d \over 2}\langle N|\bar{u}u-\bar{d}d|N\rangle = - 2.52 \;\; {\rm MeV}\; \bar{N} \tau^3 N,
\label{massdiff}
\ee
we know that 
\ba
g_a^{S'} {\bar {N}} \tau^3 N &=& ({m_u+m_d})\langle N|\bar{u}u-\bar{d}d|N\rangle \nonumber \\
&=&-2 {R+1 \over R-1 } 2.52 \;\; {\rm MeV}\; \bar{N} \tau^3 N
\ea
where $R= m_u/m_d=0.47\pm 0.04,$ is the mass ratio of the up quark to the down quark. This implies that $g_a^{S'}\simeq 13.98$ MeV.

The Lagrangian for a triplet pseudoscalar, $a^b$ for $b =1-3$, can be written as,
\be
{\cal L} \supset {g^T_a \over F_a} \partial_\mu a^b {\bar {N}} \gamma^\mu \gamma_5 {\tau^b \over 2} N
+ {g_a^{T'} \over F_a F_\pi} \pi^b a^3 \; {\bar {N}} \tau^b N,
\ee
where $g_a^T=\Delta u -\Delta d \simeq 1.264$ \cite{Bishara:2016hek} and $g_a^{T'} {\bar {N}} \tau^3 N = {m_u-m_d \over 2}\langle N|\bar{u}u-\bar{d}d|N\rangle $. 
According to Eq.~\ref{massdiff}, this implies $g_a^{T'}\simeq 2.52$ MeV.

Using the Lagrangians introduced above one can obtain the expressions for the ratios of the pseudoscalar production cross sections to that of a photon. The expression for the singlet pseudoscalar reads 
\ba
&&\frac{\left(\sigma v\right)_{\pi^-p\to a n}}{\left(\sigma v\right)_{\pi^-p\to\gamma n}}=
\left({g^S_a \over e F_a}\right)^2 {m_N^2 x^4 \over 2} \times   \\
&&\left(1-{g_{S} (x^2-y-2)(y+2) \over 4 m_N x^2 (1+y)}\right)^2 
 f_a^S\left(x,y\right),\nonumber
\ea
with $x=m_a/m_N$, $y=m_\pi/m_N$, and
\be
\begin{aligned}
&f_a^S(x,y)=\left[1-\left(\frac{x}{2+y}\right)^2\right]\left[1-\frac{x^2}{2}+\frac{y}{2}\right]^{-2}
\\
&\times\left(1+\frac{x^4-2 x^2\left(2+2y+y^2\right)}{y^2\left(2+y\right)^2}\right)^{1/2}.
\end{aligned}
\ee
Here $g_S= 2 g_a^{S'}/(g_a^{S} g_A)$.

The expression for the triplet pseudoscalar is as follows,  
\begin{align}
&\frac{\left(\sigma v\right)_{\pi^-p\to a n}}{\left(\sigma v\right)_{\pi^-p\to\gamma n}}= 
\left({g^T_a \over e F_a}\right)^2 {m_N^2 \; y^4 \over 8(1+y)^2}\, \times  \\
&\left(1+{g_{V} (x^2-y-2)(y+2) \over m_N [y^2(2+y)-x^2(2+2y+y^2)]}\right)^2\,f_a^T\left(x,y\right),\nonumber
\end{align}
with $x$ and $y$ as above and
\be
\begin{aligned}
&f_a^T(x,y)=\sqrt{1-\frac{x^2}{y^2}}\left[1-\left(\frac{x}{2+y}\right)^2\right]^{3/2}
\\
&\times\left(1-\frac{2 x^2 (1+y)}{y^2\left(2-x^2+y\right)}\right)^2.
\end{aligned}
\ee
Here $g_V= 2 g_a^{T'}/(g_a^{T} g_A)$.

\subsection{Singlet and triplet axial vectors}
Similarly, the Lagrangian for a singlet axial vector ($A_\mu$) coupling to the axial current has the following form,
\be
{\cal L} \supset {g^S_A} Q_N^A A_\mu {\bar {N}} \gamma^\mu \gamma_5 N,
\ee
where $Q_N^A$ is the axial charge of the nucleon doublet $N$. 
The Lagrangian for a triplet axial vector, $A_\mu^b$, can be written as
\be
{\cal L} \supset {g^T_A} A_\mu^b {\bar {N}} \gamma^\mu \gamma_5 {\tau^b \over 2} N
\ee

The cross section ratio for the singlet axial vector is 
\be
\frac{\left(\sigma v\right)_{\pi^-p\to A n}}{\left(\sigma v\right)_{\pi^-p\to\gamma n}}=
\left({Q_N^A g^S_A \over e}\right)^2 {y^2 \over (1+y)^2}\,f_A^S\left(x,y\right),
\ee
with $x=m_A/m_N$, $y=m_\pi/m_N$, and
\begin{align}
&f_A^S(x,y)=\left[1-\frac{x^2}{y^2}\right]\left[1-\left(\frac{x}{2+y}\right)^2\right]^2\left[1-\frac{x^2}{2+y}\right]^{-2} \nonumber
\\
&\times\left(1+\frac{x^4-2 x^2\left(2+2y+y^2\right)}{y^2\left(2+y\right)^2}\right)^{1/2}.
\end{align}

The ratio for the triplet axial vector can be written as, 
\be
\frac{\left(\sigma v\right)_{\pi^-p\to A  n}}{\left(\sigma v\right)_{\pi^-p\to\gamma n}}=
\left({g^T_A \over e}\right)^2 {y^4 \over 8 x^2 (1+y)^2}\,f_A^T\left(x,y\right),
\ee
with $x$ and $y$ given above and
\begin{widetext}
\be
\begin{aligned}
&f_A^T(x,y)=\left[1-\frac{x^2}{y^2}\right]\left[1-\frac{x^2}{2+y}\right]^{-2}
\left(1+\frac{x^4-2 x^2\left(2+2y+y^2\right)}{y^2\left(2+y\right)^2}\right)^{1/2}
\\
&\times\left(1+\frac{2 x^6 (1+y)^2-2 x^2 (2+y)^2(y^3-2y^2-8y-4)+x^4(y^4-4y^3-28y^2-40y-16)}{y^2\left(2+y\right)^4}\right).
\end{aligned}
\ee
\end{widetext}

Notice that in the case of $g_A^T$, one observes a strong enhancement of $A$ emission in the limit of $m_A\to 0$, with capture rate scaling as 
$\propto m_{A}^{-2}$ due to the nonconservation of the axial current~\cite{Dror:2017ehi,Dror:2017nsg}.

% show all refs even ones not referred to in text
%\nocite{*}
\bibliography{picap}

\end{document}